# Multi-Criteria Optimization for Image Guidance

Brian Winey* and James Balter[#]

*Department of Radiation Oncology, Massachusetts General Hospital, Harvard Medical School, Boston MA 02114

winey.brian@mgh.harvard.edu

[#]Department of Radiation Oncology, University of Michigan, Ann Arbor, MI 48109

**Abstract:**

**Purpose:** To develop a multi-criteria optimization framework for image guided radiotherapy.

**Methods:** An algorithm is proposed for a multi-criteria framework for the purpose of patient setup verification decision processes. Optimal patient setup shifts and rotations are not always straightforward, particularly for deformable or moving targets of the spine, abdomen, thorax, breast, head and neck and limbs. The algorithm relies upon dosimetric constraints and objectives to aid in the patient setup such that the patient is setup to maximize tumor dose coverage and minimize dose to organs at risk while allowing for daily clinical changes. A simple 1D model and a lung lesion are presented.

**Results:** The algorithm delivers a multi-criteria optimization framework allowing for clinical decisions to accommodate patient target variation make setup decisions less straightforward. With dosimetric considerations, optimal patient positions can be derived.

**Conclusions:** A multi-criteria framework is demonstrated to aid in the patient setup and determine the most appropriate daily position considering dosimetric goals.

## 1. Introduction

Image guided radiotherapy can deliver higher levels of confidence in accurate radiation delivery to patients. Using 2D (radiographs)[1-7] or 3D (e.g. cone beam computed tomography (CBCT))[6, 8-17] imaging at the time of treatment can aid in the accurate positioning of a patient and the geometrically confident delivery of radiation dose to the prescribed target. When treating targets that move and/or deform in the spine, abdomen, thorax, head and neck, breast and limbs, image guidance might not provide simple or straightforward patient shift calculations[2, 4, 11, 18-21]. Additionally, shifts generated from bony anatomy might not directly or accurately reflect the daily variation of a soft tissue target not attached to the bony structure[11, 18, 19]. Finally, it is well known that head and neck cancers can change size during the treatment fractionation and result in large changes in dosimetry to the target and surrounding organs at risk (OARs)[22].

To develop more accurate image guidance, several procedures have been proposed. Repeat CT and CBCT scanning is often utilized to provide a 3D alignment assessment but deformable and moving targets can not be easily aligned using static 3D images and rigid alignment techniques[8, 18, 19, 23, 24]. Additionally, the alignment is generally performed using surrogates that have variable degrees of fidelity for positioning the poorly visualized tumor and surrounding organs at risk

(typically either bony anatomy or soft tissue or a manual attempt at averaging the two)[11, 18-20]. Some groups have suggested performing daily dose calculations on the daily CT or CBCT for more informed dosimetry evaluations[23, 24]. While this concept has merit, both logistic (calculation speed and infrastructure) as well as accuracy (dependence on accurate deformable modeling and dose re-calculation/summation) limitations hinder current implementation.

Multi-criteria approaches allow experts to understand the tradeoffs inherent in their decisions, and have demonstrated usefulness in guiding treatment planning optimization[25-27]. Currently, no study has presented a multi-criteria approach to the address the tradeoffs within image guidance. To this end, we propose an algorithm that can interactively aid in the patient setup decision making process, utilizing a multi-criteria optimization interface.

## 2. Methods

### 2.1. Image Guidance Workflow

We propose the following workflow as a foundation for this algorithm:

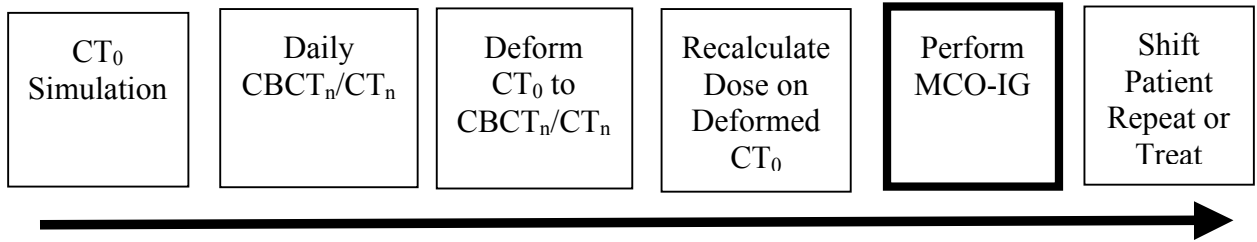

The patient simulation $CT_0$ would be transferred to a shared computer capable of performing a deformable registration. The daily $CBCT_n$ or $CT_n$ would be transferred to the shared computer and the $CT_0$ would be deformably registered to the $CBCT_n$ or $CT_n$. Dose would be recalculated on the deformed $CT_0$ and used in the MCO algorithm. The MCO algorithm produces a recommended patient shift that accounts for the dosimetric considerations of the clinical protocol. Patient would be shifted and treated or shifted and reimaged for a repeat of the process.

### 2.2. MCO for Image Guidance

Similar to MCO for treatment planning[25-27], the dose delivered to the patient is modified on a continual variable space to generate a Pareto surface. Instead of altering the fluence or beam parameters, the MCO-IG algorithm generates a Pareto from shifts and rotations of the patient. Additionally, the MCO-IG algorithm can include beam weights for non-IMRT beams or multiple plans in the optimization process. In general, the Pareto surface would be generated from the following system of equations:

$$M_i(r) = \sum_{j=1}^{n} w_{i,j} D_j(r), \; s.t. \; D_{GTV} \geq D_{\lim}^{GTV} \text{ and } D_{OAR} \leq D_{\lim}^{OAR} \qquad (1)$$

where $D_j(r)$ is a constraint or objective with weight, $w_{i,j}$, for a patient shift, $r$, and the Pareto surface defined by all possible values of $M_i(r)$ is a function of the patient shifts and the weights of the objectives and constraints summed over all $j=\{1,n\}$ objectives or constraints.

*2.3. Simple 1D Model*

For a GTV separated from an OAR by distance $\Delta x$ that can vary with each daily treatment fraction, there is a dose curve, $D(x)$, with a 80%-20% penumbra of 2 mm, Figure 1, where the ideal (planning) position of the GTV is given such that the GTV receives 99% of the dose. A simple version of Eq 1 can be reduced to a two constraint problem, dose to GTV and dose to OAR as a function of $r = x$, the setup position:

$$M_i(x) = w_{i,GTV} D_{GTV}(x) + w_{i,OAR} D_{OAR}(x) \tag{2}$$

where $x$ can vary to provide a better compromise of constraints and solutions and $w_i$ is a variable weighting between the dose to the GTV and the dose to the OAR. To simplify the display of the solution, we can substitute the weights of the equation by a ratio of the weights:

$$M_i(x) = \left(\frac{w_{i,GTV}}{w_{i,OAR}} D_{GTV}(x) + D_{OAR}(x)\right), s.t.: D_{GTV} \geq D_{\lim}^{GTV} \text{ and } D_{OAR} \leq D_{\lim}^{OAR} \tag{3}$$

The solutions of Equation 3 can be constrained by the $D_{GTV}$ such that a certain prescription level is attained or permitted to float with variable weights depending upon the clinical case.

For example, assume the GTV and OAR are separated by $\Delta x = 1, 2,$ or 3 mm's depending upon the fraction. A clinical scenario might include a lung tumor that has drifted closer to or farther from the liver and diaphragm or a prostate that has drifted closer to or farther from the rectum. From Equation 3, optimal setup positions can be determined such that the constraints or objectives are met for the given weights.

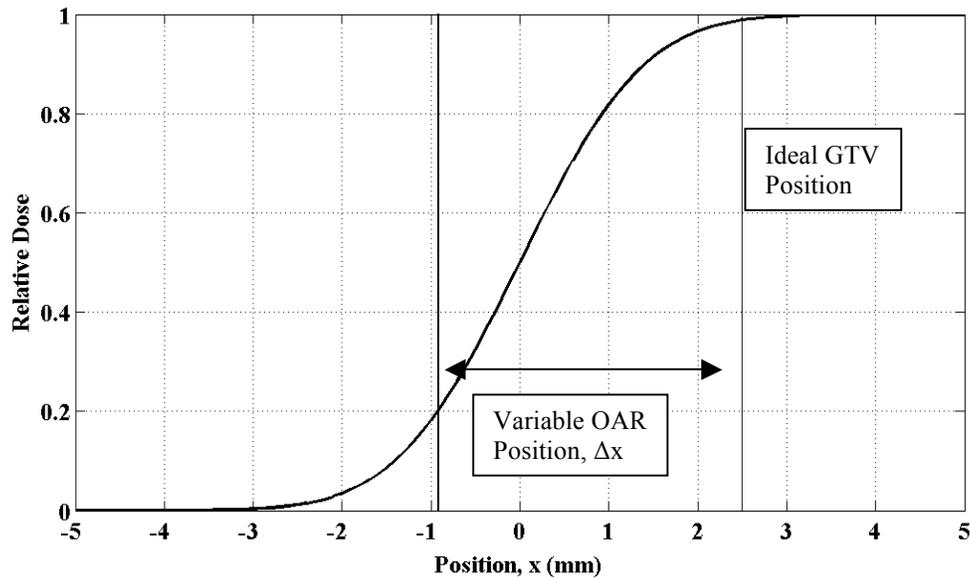

Figure 1: Dose profile as a function of position, $x$. The ideal position of the GTV is noted as the location where the GTV receives 99% of the dose.

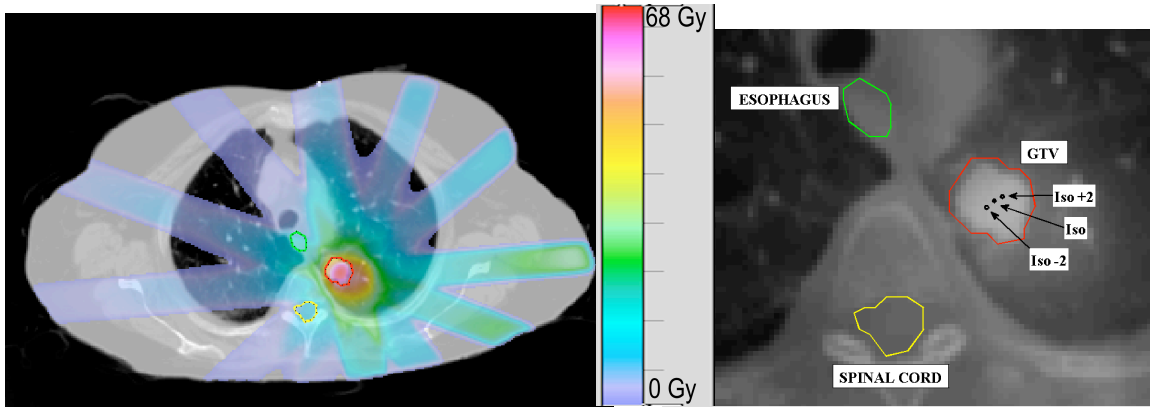

Figure 2. A lung lesion with nearby OAR's of the spinal cord and the esophagus. Simulated treatment position variations of +/- 2 mm are demonstrated on the enlarged image. The dose in (a) is normalized to the 70% isodose line for a 48 Gy prescription.

*2.4. 2D Lung Case*

For a lung lesion demonstrated in Figure 2, there are two OAR's in the vicinity of the GTV: the esophagus and the spinal cord. Small shifts in the position of the patient, relative to the planned dosimetric isocenter can result in large differences in the dose delivered to the OAR's and GTV by pure geometric shifts of the OAR or GTV within the dose cube.

In the 2D example, equation 3 will change such that $x \rightarrow x,y$ without consideration of rotations. As a first order approximation over small deviations of position of the patient, the dose cube will be assumed to remain constant and doses to the OAR's and GTV will change only as a function of the shifts in $x,y$.

For the purpose of simulating possible shifts of the GTV from the time of simulation to the treatment delivery, two relative shifts of the GTV with respect to the OAR's are displayed in figure 2b, one of -2mm toward the OAR's and one of +2mm away from the OAR's.

In the clinical scenario of the lung lesion, the spinal cord might have a tight dose gradient nearby and the physician would prefer less dose to the OAR while compromising the inferior medial edge of the GTV. Or, if the distance increased between simulation and the treatment fraction, the clinician might prefer to deliver a higher isodose to the PTV. In either case the relative weights between the $D_{PTV}$ and the $D_{OAR}$ would reflect the clinical decision at the time of treatment and an optimal setup position would be determined.

For the case displayed in Figure 2, the treatment plan was developed in CMS (Elekta, St Louis, MO USA) using 12 conformal beams. The prescription dose was 48 Gy in 4 fractions and was prescribed to the 70% isodose line. The high dose region is not centered on the GTV due to the proximity of the OAR's and a clinical decision to pull the high dose region away from the OAR's. The dose distribution of this patient specific example will influence the results of the MCO calculations.

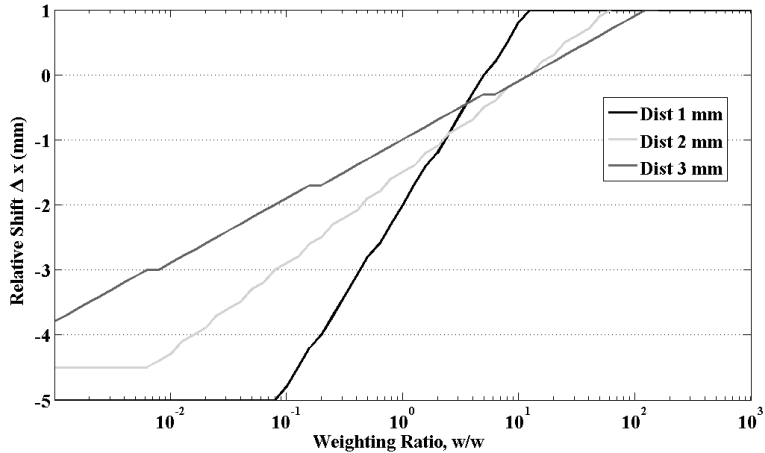

Figure 3: The optimal shifts for the varying ratio of the weights given in Equation 3. The shift of zero corresponds to the GTV receiving 99% of the dose. The asymptotic regions correspond to the GTV receiving 100% of the dose (the right side) and the OAR being completely avoided (left side).

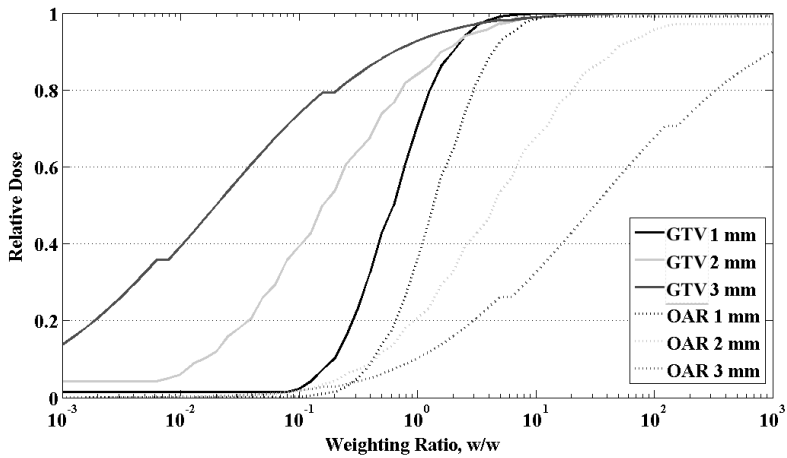

Figure 4: Dose to the GTV and OAR as a function of the weighting ratio in Equation 3 and separation between the GTV and OAR. The 1 mm case is given by the black solid and dotted black line for the GTV and OAR, respectively. The range of optimal cases is limited by the small separation of the GTV and OAR. Larger separations provide greater flexibility of optimal setup positions and greater choices for the clinician.

## 3. Results

The simple 1D model solution of the equation is presented for the 1,2, and 3 mm separations in Figures 3 and 4 as a function of the relative offset position compared to a perfect alignment of the GTV to the planning position and as a function of the resulting dose to the GTV and OAR's. A negative shift is equal to moving the GTV/OAR to the left in the dose plot of Figure 1 such that the doses to the GTV and OAR decrease. The clinician can make a more educated decision regarding the optimal setup position of the GTV given this information.

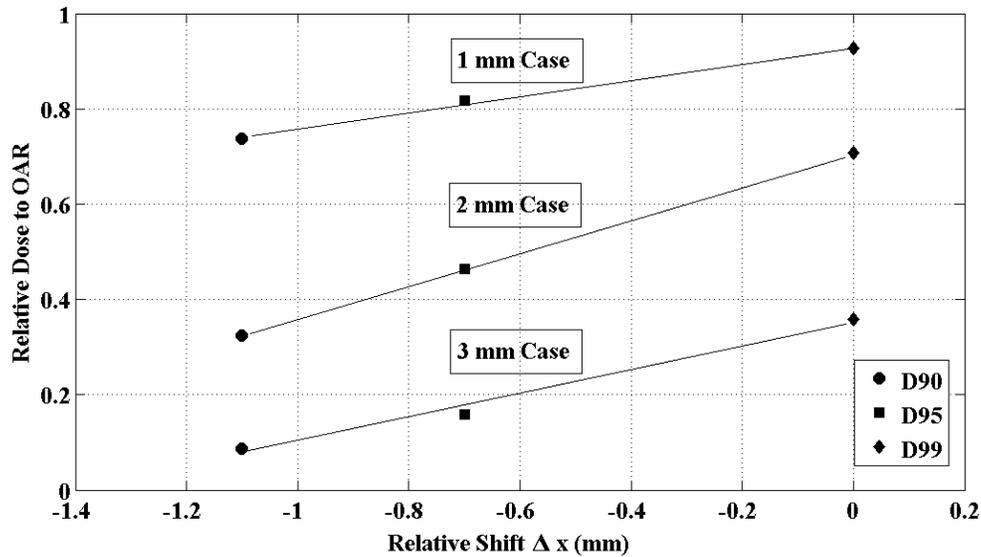

Figure 5. The optimal shift of the patient is displayed versus the respective dose to the OAR for a given dose constraint to the GTV. The solution is straightforward in 1 dimension as the algorithm pushes the GTV to the left (negative) until the dose constraint is met.

In Figure 3, the slope of the line for the 3 mm separation is less than the 1 and 2 mm separation cases due to the reduced impact of shifts to Equation 3 because the dose to OAR and thus has less impact on the optimization equation. Figure 3 displays the resulting doses to the GTV and OAR for the variable weighting ratio and for the three separation scenarios. As can be seen from the plot, the smallest separation results in a limited range of weighting factors over which the optimization can be performed. When the separation increases to 3 mm, the range of weighting values increases, providing greater flexibility in solutions.

Additionally, Figure 5 presents the optimal shifts for constraints of dose to the GTV such that the minimum dose is 99%, 95%, and 90%. As can be seen, in this simple 1D model, the shifts determined by the MCO simply force the GTV to the left until the dose limit is achieved. Since the ideal position of the GTV is such that it receives 99% of the dose, the D99 points all produce a 0 magnitude shift. The other shifts correspond to the distance of the dose limit along of the dose curve from the ideal position. For more complicated combinations of translations, rotations and dose distributions, the optimal patient shift would not be straightforward.

For the 2D lung GTV example of Figure 2, the dose is constrained by the OAR's in the vicinity of the GTV. As an example of possible differences between simulation and treatment, two relative offsets of the GTV with respect to the OAR's and patient were included, 2 mm toward the OAR's and 2 mm away from the OAR's.

Figure 6 displays the average doses to the OAR's ($D_{OAR}(x,y)$) and the minimum dose to the GTV ($D_{GTV}(x,y)$) as functions of various positions of the isocenter. Figure 6a is for the planned relative positions of the GTV and OAR's and (b) and (c) are for the relative shifts of the GTV toward and away from the OAR's, respectively.

The local maxima of the 2D version of equation 3 are also provided as large "+" symbols on figure 6a and it can be seen that equation 3 maximizes on a line along the minimum OAR doses and maximal GTV minimum dose. Also, figure 6a displays the global maxima for relative weighting factors of 1,1/5 and 1/10 as the large square, circle, and triangle, respectively. The shifts suggested by the optimal solutions of equation 3 for the three different weighting ratios are displayed in Figure 7 as black dots.

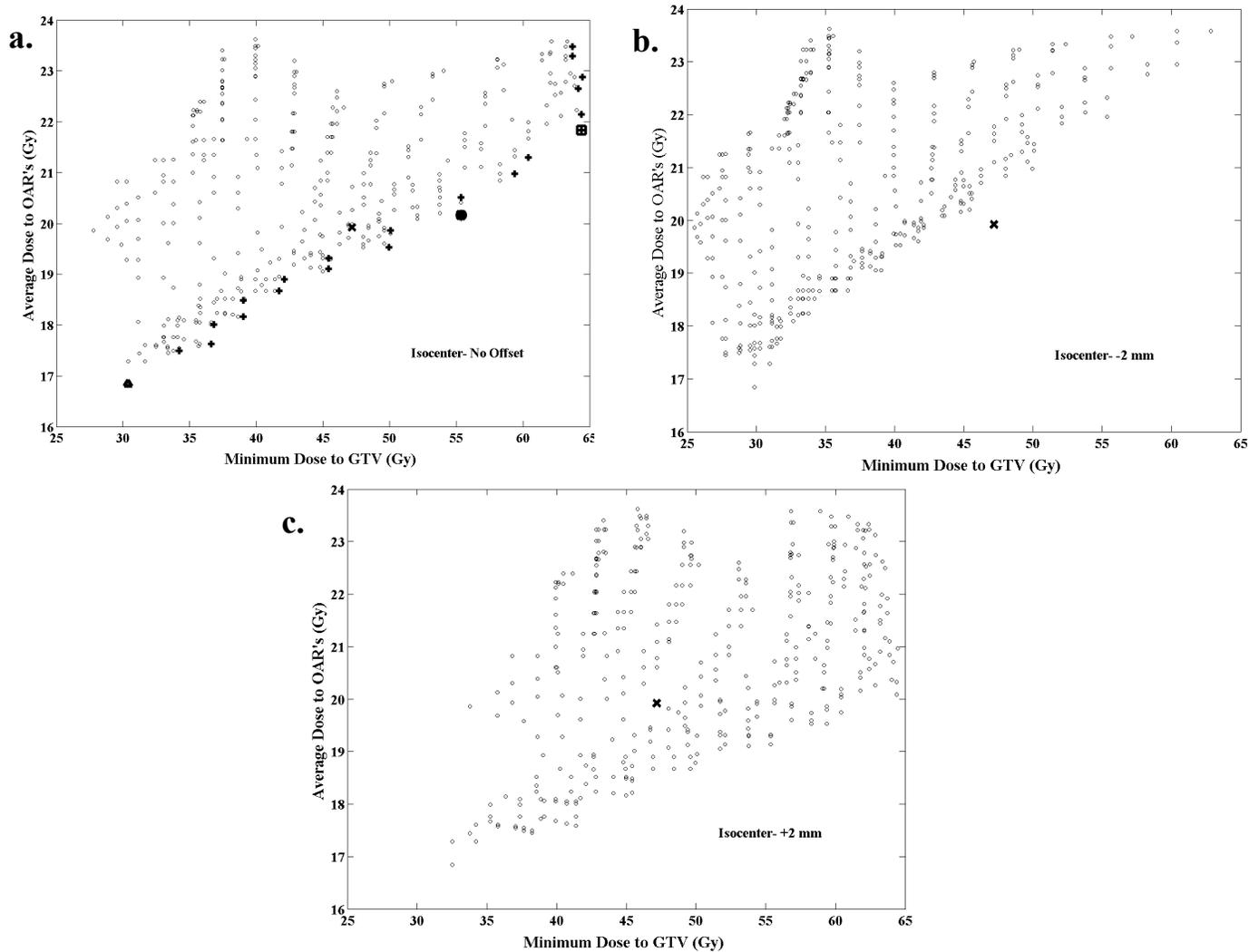

Figure 6. The dosimetric results ($D_{GTV}(x,y)$ and $D_{OAR}(x,y)$) for the GTV and the OAR's when simulating an ideal reproducible setup (a), an offset of -2 mm from the planned position (b), and an offset of +2 mm from the planned position (c). The large X corresponds to planned dose distribution with perfect setup. The large + symbols in (a) correspond to the local maxima of the equation 3, and the square, circle and triangle correspond to the global maxima of equation three for relative weights of 1, 1/5, and 1/10, respectively. The prescription dose was 48 Gy at the 70% isodose level.

In the case of the patient specific lung example, the dose distribution is not ideally centered on the GTV due to the planning constraints of the OAR's in the vicinity of the GTV and thus the recommended shifts of the patients tend to shift the GTV more into the higher dose region of the dose distribution when the dose to the OAR's are not heavily weighted.

In the planned isocenter case, figure 6a, it can be seen that the optimal patient position might not be the planned isocentric setup. Rather, the dose to the GTV could have been increased by almost 10 Gy for a modest increase of 0.5 Gy to the OAR's (Ratio of 1/5 position in figure 7 and large circle on figure 6a) with a small shift of the patient position from the planned position. In other words, assuming the patient position on the day of treatment exactly matched the simulation position, the dose to the GTV could have been increased nearly 10 Gy with only an extra 0.5 Gy to the OAR's with a small shift in the patient position. Conversely, the prescription isodose could have been raised to 85% to achieve the same dosimetric coverage of the GTV and greatly reducing the dose to the healthy tissue.

In the case that the GTV shifts from the planned position relative to the OAR's (figures 6b and 6c), it is evident that compromises of planned dose and OAR dose need to be made when the relative distance between the GTV and OAR's decreases. When the distance increases, greater dose can be delivered to the GTV and less dose to the OAR's for a given shift in the patient position.

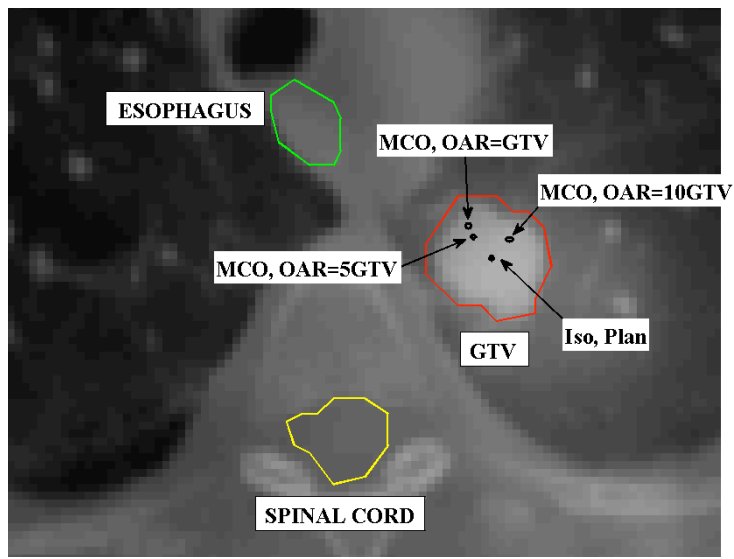

Figure 7. A display of the effective offsets as determined from the 2D version of equation 3 for three different relative weighting factors, 1,1/5, and 1/10 of GTV/OAR doses.

## 4. Conclusions

Image guidance in radiotherapy is not a simple process for some patients, particularly when the target moves within the patient or the daily patient setup introduces patient deformations. In the cases when the optimal patient position is not straightforward, MCO-IG might provide a more extensive consideration of the possible solutions that incorporate the dosimetric aims of the clinical protocol.

The simple 1D solution presented in this paper can be easily solved without MCO but it serves to demonstrate the scenarios where MCO might aid the determination of the optimal patient position for a given treatment fraction. Given a variable of GTV and OAR separation, the model presents multiple possible solutions from which the clinician can select the preferred patient shift.

A 2D example demonstrates that for a patient specific dose distribution, a planned patient position might not deliver the ideal dose to the GTV and OAR's and instead a small shift in the patient position from the planned position can increase the dose to the GTV for equivalent OAR doses. Additionally, relative shifts in the GTV with respect to the OAR's can require compromises of doses to the GTV and OAR's if the distance between the high dose regions and the OAR's decreases. Or, GTV doses can be increased without an increase of dose to OAR's when the distance between the GTV and OAR increases.

A more complicated case would include more than two dose constraints (i.e. $D_{MAX}$ to the cord, gEUD, NTCP) and all six degrees of patient position freedom to form a muli-dimensional Pareto surface. Additionally, the MCO algorithm could include beam weights as a variable for simple 3D conformal fields or multiple plans for intensity modulated deliveries. The ultimate goal is to deliver the optimal treatment plan considering daily patient variations given a set of dosimetric goals and MCO is a possible solution for achieving this aim.

**Acknowledgements:**

We would like to thank David Craft, Ph.D., for his comments regarding multi-criteria optimization and its application to image guidance problems.